\documentclass[showpacs,twocolumn,floatfix,prb]{revtex4}

 \usepackage{graphicx}
 \usepackage{hyperref}

 \newcommand{\bE}{\mathbf{E}}
 \newcommand{\br}{\mathbf{r}}
 \newcommand{\bez}{\mathbf{e}_z}
\begin{document}

 \title{Low crosstalk intersections of CCWs composed by mono-mode cavities}
 \author{Weiqiang Ding}
 \email{wqding@hit.edu.cn}
 \author{Lixue Chen}
 \author{Wenhui Li}
 \author{Shutian Liu}
 \address{Applied Physics Department, Harbin Institute of Technology, Harbin, 150001, PRC}

 \begin{abstract}
 Low crosstalk intersections formed by two crossing coupled cavity waveguides (CCWs),
 which are formed by series of \textit{mono-mode} cavities
 in a two dimensional photonic crystal structure,
 are investigated. Although the individual cavities are \textit{mono-mode}, the modes of
 the properly designed
 CCWs (which are called \textit{supermodes} in this paper) may be dipole-like, and
 antisymmetry along the axis of the other perpendicular CCW,
 which ensures the low
 crosstalk of the intersections. According to the results of coupled mode analysis,
 we divide the intersections into two kinds based on the number of cavities of each CCW,
 and the transmission and
 crosstalk spectra of both kinds are numerically obtained using finite difference time domain (FDTD) method.
 The results show that crosstalk power is lower than $-30$dB. We also discuss the
 factors that affect the performance of the intersections.

 \end{abstract}
 %%42.79.Gn: Optical waveguides and couplers;
 %%42.70.Qs Photonic bandgap materials
 \maketitle

 %%%%%%%%%%%%%%%%%%%%%%%%%%%%%%%%%%%%%% body %%%%%%%%%%%%%%%%%%%%%%%%%%%%%%%%%%%%%%%%%%

 \section{Introduction}

 Low crosstalk intersections of optical waveguides are key components in compact and
 high dense optical integrated
 circuits. Usually, the intersection of  two crossing waveguides should be designed and
 optimized properly, or
 large crosstalk and reflection may be appear. For the intersections formed by two conventional
 planar waveguides, mode expander structures may be
 used to decrease crosstalk and reflection (intersection \#0).\cite{cros0:jjap04}
 For the intersections of line defect
 waveguides in photonic crystals (PCs),\cite{PC:EYa,PC:SJohn} several designs
 have been proposed:\cite{cros1:StJohn,cros2:SLan,cros3:OL04,cros3:ptl05}
 Previously, the two crossing PC line defect waveguides are
 coupled to a common resonator (a point defect) (intersection \#1), and the proper symmetry of
 the resonator's mode ensures the low crosstalk between the two crossing waveguides.\cite{cros1:StJohn}
 Subsequently, this idea is used to design the crossing of two
 \textit{modified} planar waveguides  (with cavities or cuts in them).\cite{cros1:StJohn,cros0:jlt99}
 However, due to the single mode property of
 the resonator, only one operation frequency (the frequency can be tuned by the resonator)
 is demonstrated to be low crosstalk in
 Ref.\onlinecite{cros1:StJohn}.
 Recently, S. Lan and H. Ishikawa proposed a broad
 bandwidth operation of line defect PC waveguide intersections by replacing the single
 resonator of intersection \#1 with a coupled cavity structure (intersection \#2).\cite{cros2:SLan}
 Due to the wide broadband transmission band of the CCW,\cite{ccow:yariv,ccow:prl} broad
 bandwidth operation of low crosstalk intersection is numerically
 verified in a triangle lattice PC structure, \cite{cros2:SLan}
 which is also valid for ultrashort (about 500-fs) pulse transmission for the broadband property.
 More recently, a new design and optimization of line defect PC waveguide crossing
 based on Wannier basis is investigated in a square lattice
 (intersection \#3).\cite{cros3:OL04,cros3:ptl05,wannier} The bandwidth of the optimized
 crossing is very high ($\delta\lambda/\lambda\sim2\%$), and
 crosstalk is very low ($-40$dB).\cite{cros3:ptl05}

 Apart from the line defect waveguide,
 another important kind of waveguide in PC is the coupled cavity waveguide
 (CCW).\cite{ccow:prl,ccow:yariv} Although many novel optical functional elements based on
 CCW structure have been proposed, such as the optical power
 splitter,\cite{ccow:spliter:apl,ccow:spliter:apl03} band drop filter,\cite{ccow:banddrop}
 nonlinear enhancement,\cite{ccow:nl:oe,ccow:nl:prb} optical delay
 lines\cite{ccow:delay,ccow:hop} and optical bistable switching and limiting,\cite{ccow:ob:conti,ccow:OL:conti}
 the intersections of CCWs have not been
 investigated deeply. In fact, the intersection \#2 of \textit{hybrid waveguides} structure in
 Ref. \onlinecite{cros2:SLan} is a design for two crossing CCWs.

 Generally speaking, all the intersections mentioned above are all
 based on the orthogonality of the field pattern of one waveguide in the
 intersection area with respect to the other crossing waveguide.
 Therefore, the cavities of intersection \#1 should be
 multi-pole mode, and a mono-pole mode is not suitable for low
 crosstalk intersection.\cite{cros1:StJohn,cros2:SLan} When a CCW is used, the intersection
 using the mono-pole mode of the cavity is also impossible.\cite{cros2:SLan}

 In this paper, we report a new mechanism of low crosstalk
 intersection of two CCWs, which are formed by
 sequences of \textit{mono-mode} cavities in a square lattice
 PC structure. Although the individual cavities are mono-mode, the \textit{supermodes} (result from
 the strong coupling between cavities) of the CCW
 may be dipole-like, which ensures the low crosstalk between the two
 intersecting CCWs, and the working frequencies can be tuned easily.

 \section{Mode patterns of the supermodes}
 It's reported that the eigenfrequency spectra of CCW structures follow into two
 different shapes:\cite{coupl12D,ultrashort} One is a continuous band and the other is a
 series of $N$ (the number of cavities) discrete modes.\cite{supermode:dwq} And
 the shapes of the
 spectrum depend on the coupling strength between cavities.\cite{coupl12D,ultrashort}

 For the continuous spectrum modes, the coupling between neighboring cavities is weak, and
 the tight binding theory gives a
 complete description.\cite{ccow:yariv,ccow:prl}  While for the discrete
 modes, the coupling between neighboring cavities is strong, and
 we surprisingly find that a coupled mode
 theory may be used to predict the eigenfrequencies, mode profiles
 and as well as the quality factors of each supermodes.\cite{dwq:oc05,supermode:dwq}

 In the coupled mode theory, the electric field of the overall CCW is expressed by the
 superposition of the individual cavity modes with the superposition coefficients of
 arbitrary complex numbers of $A_n$: \cite{supermode:dwq}
 \begin{equation}\label{eq:sumEcmt}
 \bE_{\omega}(\br)=\sum_{n=1}^{N} A_n \bE_{\Omega}(\br-nR\bez)
 \end{equation}
 Where $N$ is the total number of cavities, and $\Omega$ and $\omega$ are the eigenfrequencies
 of an individual cavity and that of the coupled system. $\bE_{\Omega}$ and $\bE_{\omega}$ are
 the eigenmodes of an individual cavity and that of the CCW. $R$ and $\bez$ are the
 distance between two neighboring cavities and the unit vector of the alignment
 direction of the cavities, respectively. Substituting
 Eq. (\ref{eq:sumEcmt}) into the simplified Maxwell's equation. Then operate the
 obtained equation using $\int d\br\bE(\br-mR\bez)\cdot$, one can obtain a group of
 coupled linear equations about the coefficients of $A_n$.
 Solving the equation group, one obtain
 $N$ allowed supermodes for the CCW system. For the $L$th ($L=1,2\cdots,N$) mode, the
 superposition coefficients $A^L_n$ are:
 \begin{eqnarray}
 A_n^L &=& A^L_0\sin(n\theta^L), \quad n=1,\cdots,N \label{eq:AnL}\\
 \theta^L &=& \frac{L\pi}{N+1},\quad L=1,\cdots,N
 \end{eqnarray}
 Where $A^L_0$ is a normalized constant. For the details of the coupled mode
 theory, the readers are referred to Ref. \onlinecite{supermode:dwq}.

 \begin{figure}[htb]
 \centering
 \includegraphics[width=0.4\textwidth]{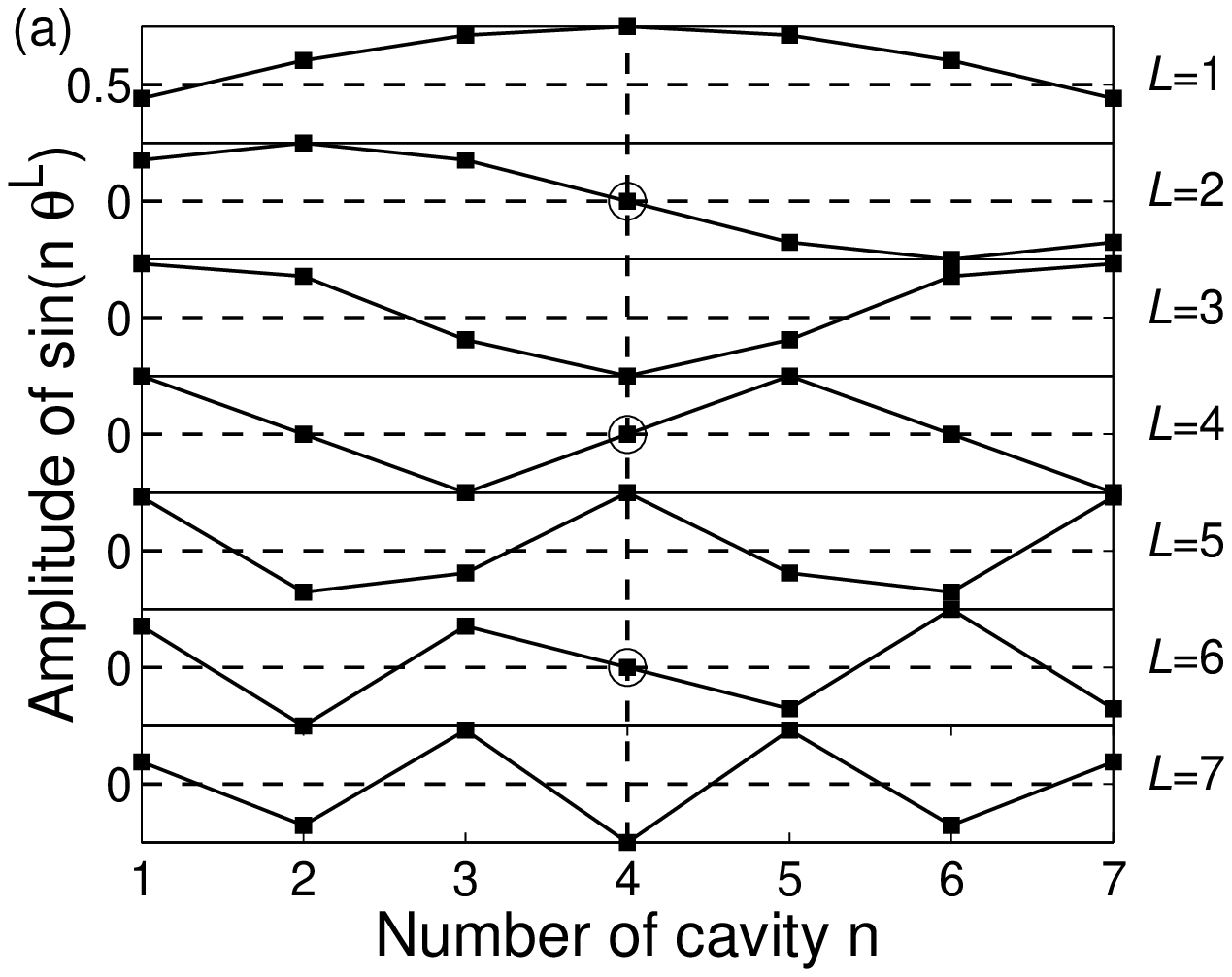}
 \includegraphics[width=0.4\textwidth]{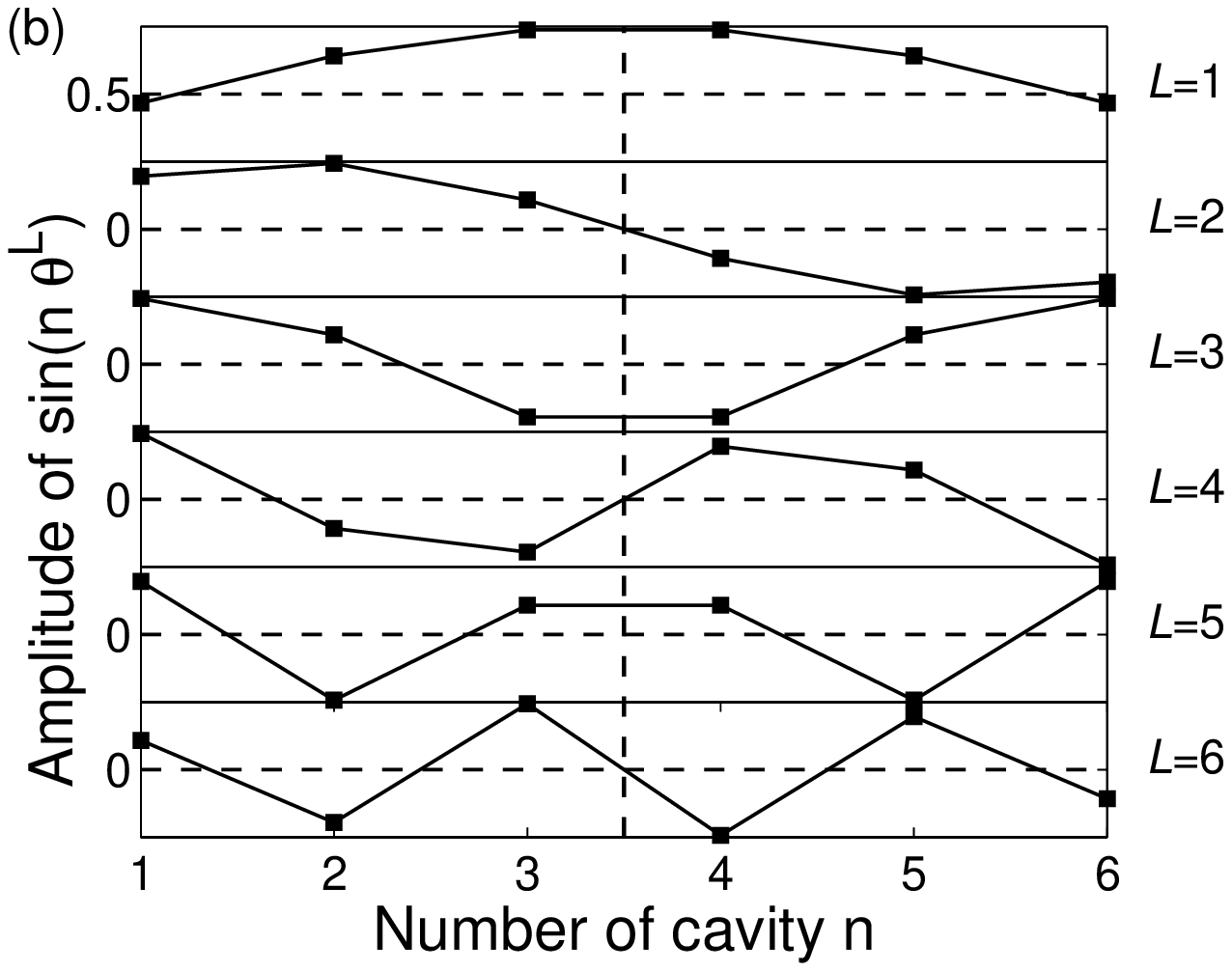}
 \caption{(a) Values of $\sin(n\theta^L)$ for the case of $N=7$,
 which is proportional to
 the superposition coefficients $A_n^L$ (Eq. (\ref{eq:AnL})). The
 distribution of $A_n^L$ is antisymmetry about the dashed vertical
 line. (b) The same as (a) except for $N=6$.}\label{fig:anL}
 \end{figure}

 One of the most important results of the theory is the determination of the superposition
 coefficients of Eq.(\ref{eq:AnL}), which measure the contribution of each cavity
 to the $L$th supermode.
 For the case of $N$ being an odd integer, i.e., $N=2m+1$ with $m$ an integer,
 one can find
 that the amplitude of the central ($(m+1)$th, $n=m+1$) cavity of the $L$th supermodes
 are $A_{m+1}^L=A_0^L\sin(L\pi/2)$. Therefore, for the case of $L=2,4,\cdots,2m$,
 $A_{m+1}^L=0$. More importantly, the amplitudes of the $l$th ($l=1,2,\cdots,m$) cavity modes to
 both sides of the $(m+1)$th
 cavity are antisymmetry, i.e, $A^L_{m+1+l}=-A^L_{m+1-l}$. For the other cases of
 $L=1,3,\cdots,2m+1$, $|A_{m+1}^L|=1$. Fig. \ref{fig:anL}(a) shows the
 values of $A_n^L$ for the case of $N=7$. Clearly, for the $4$th cavity, the
 coefficients of $A_4^L$ are all zero for $L=2,4,6$, and more importantly,
 the overall distribution of
 the coefficients are antisymmetry about the center of the CCW (shown by the dashed line
 in Fig. \ref{fig:anL}).

 For the case of $N$ being an even integer, i.e., $N=2m$ with $m$ an
 integer, the coefficients of the central cavities (the $m$th and $(m+1)$th cavities) are not
 zero. However, the coefficients are also antisymmetry about the center of the
 structure for the cases of $L=2,4,\cdots,2m$, and symmetry for the cases of
 $L=1,3,\cdots,2m-1$. Fig. \ref{fig:anL}(b) shows the
 coefficients for the case of $N=6$, and the symmetrical axis is
 shown by the dashed line.

 From the analysis and theoretical results presented above, one can find that although the
 field pattern of an individual cavity is mono-pole symmetry, half of the supermodes
 of the CCW may be dipole-like, of which the mode patterns are antisymmetry about the
 center of CCW. If another CCW is set
 perpendicular to the CCW, and the intersection area is set at the
 center of the CCW, it is possible to achieve low crosstalk intersection for the
 for the $m$ supermodes of $L=2,4,\cdots,2m$.

 \begin{figure}[hbt]
 \centering
 \includegraphics[width=0.4\textwidth]{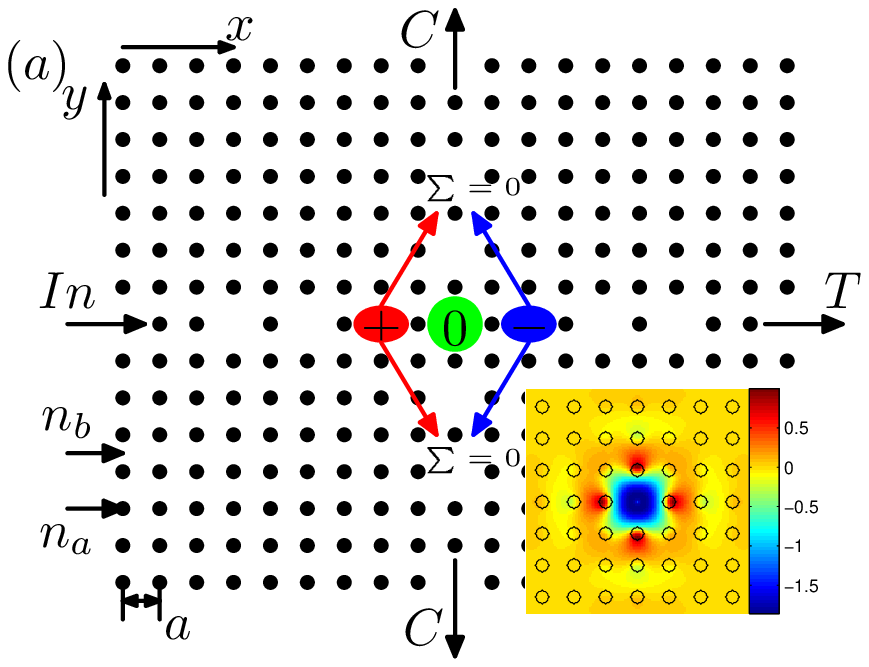}
 \includegraphics[width=0.4\textwidth]{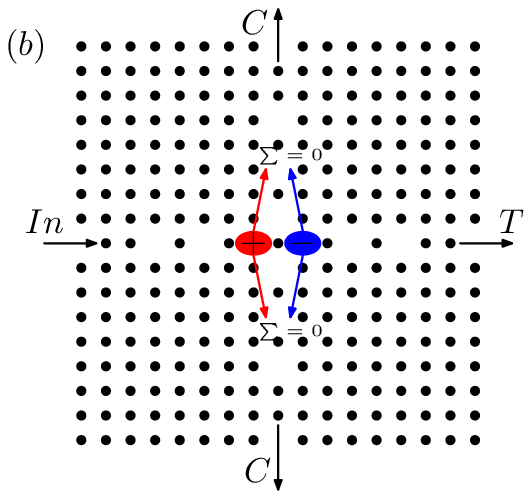}
 \caption{Schematic structure of intersections formed by two CCWs
 (along $x$ (10) and $y$ (01) directions respectively) in a square lattice
 photonic crystal (PC). The PC is formed by arranging circular
 cylinders according to square lattice in vacuum ($n_b=1$). The radii of the
 cylinders are $0.2a$ ($a$ is the lattice constant)
 and refractive index is $n_a=3.4$.
 The CCWs are fabricated by removing every
 another cylinders completely. The colorful ellipses and arrows at the intersections
 sketch the mechanism of low crosstalk. The inset of (a) shows the mono-mode pattern of an individual
 cavity at the frequency of $0.3784(2\pi c/a)$. (a) Type-A intersection,
 in which the number of cavity along $x$ direction ($N_x$) and $y$
 direction ($N_y$) are both odd integers.
 (b) Type-B intersection, in which the number of cavity along $x$ direction ($N_x$) and $y$
 direction ($N_y$) are both even integers.}\label{fig:schem}
 \end{figure}

 \section{Numerical results and discussion}

 As an example of illustration, we design intersections of two
 CCWs in a frequently used square lattice PC structure, which is formed by dielectric
 cylinders in vacuum ($n_b=1$), as shown in Fig.\ref{fig:schem}.
 The refractive indexes of cylinders are $n_a=3.4$, and the radii of
 the cylinders are $0.2a$, with $a$ the lattice constant. This PC structure opens a large
 band gap of $(0.29\sim0.42)(2\pi c/a)$ for the TM polarization (electric field vector
 parallel to the axes of the cylinders). An individual cavity is
 fabricated by removing a cylinder completely. According to the above results of coupled mode
 theory, we divide the intersections into two types, as shown in
 Fig. \ref{fig:schem}, i.e,. type-A (Fig.\ref{fig:schem}(a))
 and type-B (Fig.\ref{fig:schem}(b)).

 For the type-A intersection, the numbers of cavities along both $x$ ($N_x$)
 and $y$ ($N_y$) directions
 are odd. Without loss of generality, we set $N_x=7$ and $N_y=5$,
 respectively. And the two central cavities (the $4$th along $x$ direction and the $3$th along
 $y$ direction) are overlapped.
 The single cavity
 supports a monopole resonate state at the frequency of $\omega=0.3784(2\pi c/a)$,
 as shown in the inset of Fig. \ref{fig:schem}(a).
 For the type-B intersection, the $N_x$ and $N_y$ are both even
 integers, as shown in Fig. \ref{fig:schem}(b).

 According to the analysis in Ref.\onlinecite{cros1:StJohn} and \onlinecite{cros2:SLan},
 the cavities, which support mono-modes only, are
 not suitable for low crosstalk operation. However, for the supermodes
 of the CCWs, low crosstalk is achieved successfully in the two structures of Fig. \ref{fig:schem}.
 Using the finite difference time domain (FDTD) method,\cite{fdtd:taflove} the normalized
 spectrum of transmission and crosstalk are numerically derived. For the FDTD simulation,
 $20$ spatial grids are divided in a lattice constant $a$,
 and perfect matched layers (PML) absorbing boundary conditions (ABC) are set
 around the structures.

\begin{figure}[hbt]
 \centering
 \includegraphics[width=0.45\textwidth]{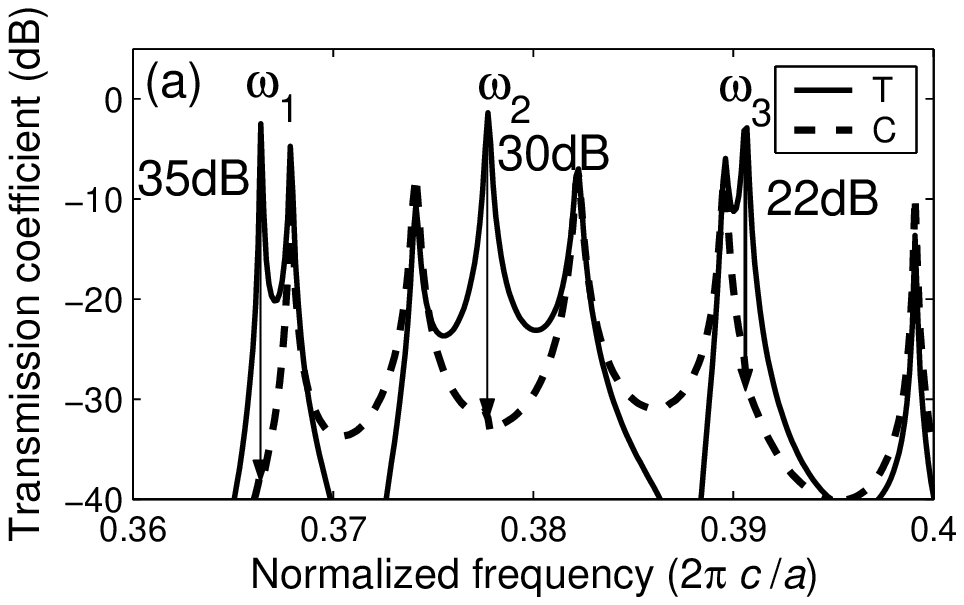}
 \includegraphics[width=0.45\textwidth]{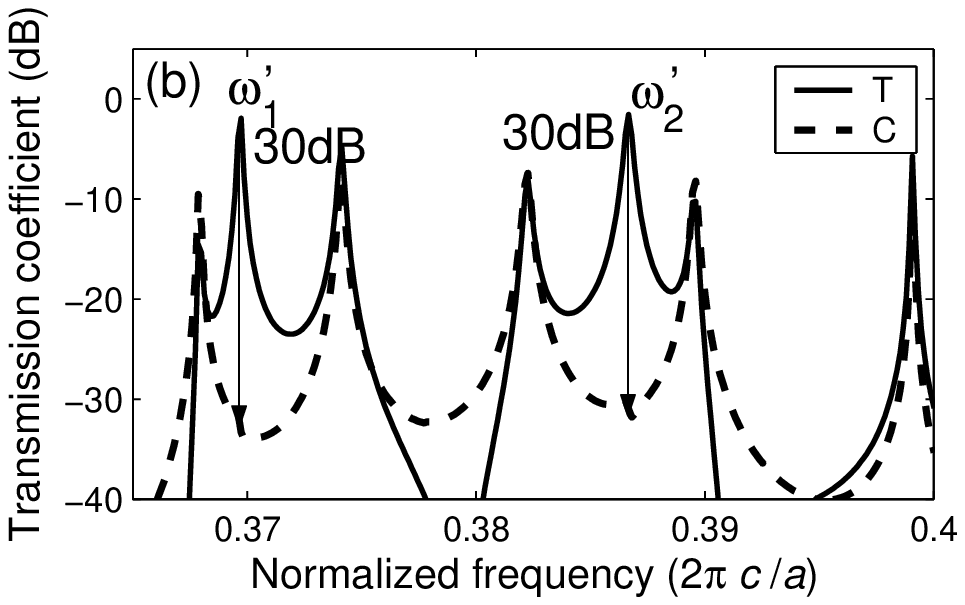}
 \caption{Transmission and crosstalk spectra of the type-A intersection shown in
 Fig. \ref{fig:schem}(a). (a) Incident along $x$ direction. The three low crosstalk
 frequencies are $\omega_1=0.3664(2\pi c/a)$, $\omega_2=0.3777(2\pi c/a)$ and
 $\omega_3=0.3906(2\pi c/a)$, respectively. (b) Incident along $y$
 direction. The two frequencies of low crosstalk are $\omega_1^{'}=0.3697(2\pi c/a)$ and
 $\omega_2^{'}=0.3697(2\pi c/a)$, respectively.}\label{fig:Ta}
 \end{figure}

 Fig. \ref{fig:Ta} shows the normalized transmission (solid lines) and
 crosstalk (dashed lines) power of
 the type-A intersection (shown in Fig.\ref{fig:schem}(a)) when incident along $x$ direction
 (Fig.\ref{fig:Ta}(a)), and $y$ direction (Fig.\ref{fig:Ta}(b)), respectively.
 Clearly, for the $x$ direction incident, the three frequencies
 of $\omega_1=0.3664(2\pi c/a)$, $\omega_2=0.3777(2\pi c/a)$ and
 $\omega_3=0.3906(2\pi c/a)$ transmit
 through the intersection with corresponding crosstalk of
 about $-35$dB, $-30$dB and $-22$dB, respectively.
 When the signal incident along $y$ direction, there are two frequencies of
 $\omega_1^{'}=0.3697(2\pi c/a)$ and
 $\omega_2^{'}=0.3697(2\pi c/a)$ transmit with a crosstalk ratios of both about $-30$dB.

\begin{figure}[hbt]
 \centering
 \includegraphics[width=0.45\textwidth]{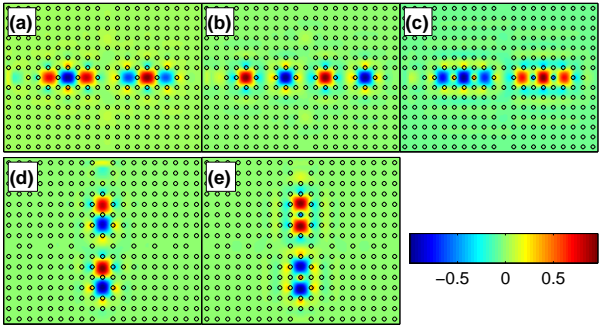}
 \caption{(Color online). Snapshots of electric field intensity distribution at stable states
 in the FDTD simulation processes.
 (a), (b) and (c) correspond to the frequencies of $\omega_1$, $\omega_2$ and $\omega_3$
 in Fig. \ref{fig:Ta}(a) respectively. (d) and (e) correspond to the frequencies
 of $\omega_1^{'}$ and $\omega_2^{'}$ in Fig. \ref{fig:Ta}(b), respectively.} \label{fig:Eina}
 \end{figure}

 Fig. \ref{fig:Eina} shows the snapshots of the electric field distribution
 at steady state of the low crosstalk frequencies. Fig. \ref{fig:Eina}(a), (b) and (c)
 correspond to the
 stable states of $\omega_1$, $\omega_2$ and $\omega_3$ in Fig. \ref{fig:Ta}(a). Fig.
 \ref{fig:Eina}(d) and (e) correspond to the $\omega_1^{'}$ and $\omega_2^{'}$ in
 Fig. \ref{fig:Ta}(b).

\begin{figure}[hbt]
 \centering
 \includegraphics[width=0.45\textwidth]{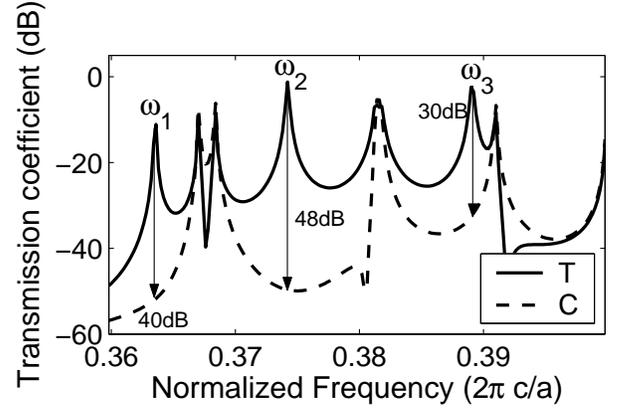}
 \caption{Transmission and crosstalk spectra of the type-B
 intersection shown in Fig. \ref{fig:schem}(b). The three low crosstalk
 frequencies are $\omega_1=0.3698(2\pi c/a)$, $\omega_2=0.3812(2\pi c/a)$
 and $\omega_2=0.3974(2\pi c/a)$, respectively.}
 \label{fig:Tb}
 \end{figure}

\begin{figure}[hbt]
 \centering
 \includegraphics[width=0.45\textwidth]{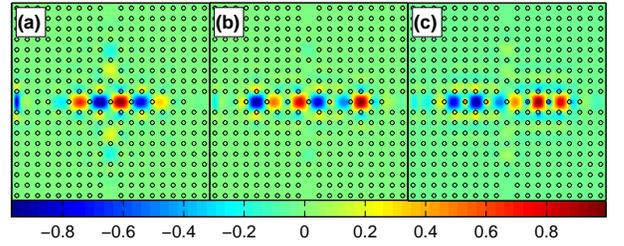}
 \caption{(Color online). Snapshots of electric field intensity distribution at stable states
 in the FDTD simulation processes.
 (a), (b) and (c) correspond to the frequencies of $\omega_1$, $\omega_2$ and $\omega_3$
 in Fig. \ref{fig:Tb} respectively.}
 \label{fig:Einb}
 \end{figure}

 For the type-B intersection, similar results are obtained. The
 transmission and crosstalk spectra are shown in Fig. \ref{fig:Tb}.
 For the three frequencies of $\omega_1=0.3698(2\pi c/a)$, $\omega_2=0.3812(2\pi c/a)$
 and $\omega_2=0.3974(2\pi c/a)$, the crosstalk ratios are
 about $-40$dB, $-48$dB and $-30$dB, respectively.
 Fig. \ref{fig:Einb} (a), (b) and (c) show the electric field intensity
 distribution of the three frequencies of $\omega_1$, $\omega_2$ and $\omega_3$ respectively.

 From the results of coupled mode analysis, we have predicted that for the
 cavity number of $2m$ and $2m+1$, there are $m$ frequencies of which
 the mode patterns satisfy the conditions of low crosstalk.
 The numerical results of Fig. \ref{fig:Ta}, \ref{fig:Eina},
 \ref{fig:Tb} and \ref{fig:Einb} confirm our prediction. For the
 case of $N_x=6,7$ and $N_y=5$, the number of low crosstalk frequencies are
 $3$ and $2$. From the electric field distribution of Fig. \ref{fig:Eina}
 and Fig. \ref{fig:Einb}, the low crosstalk mechanisms can be understood more clearly.
 One can see that near the intersection, there is a $\pi$ phase
 difference for the cavity modes to both sides of the crossing CCW, such as the 3rd and 5th cavities in
 Fig. \ref{fig:Eina} (a-c), and the 3rd and 4th cavities in
 Fig. \ref{fig:Einb}. Therefore, the tunneling signals of them
 interfere destructively in the perpendicular CCWs, and this is the
 physical mechanism of low crosstalk.

 According to the above discussion and numerical results, one can
 find that the number of cavity $N$ ($N_x$ and $N_y$) affects the performance of
 the intersection greatly. Only the type-A and
 type-B structures can eliminate the crosstalk efficiently, i.e., the $N_x$ and $N_y$
 must be both odd integers or both be even integers. If one of
 $N_x$ and $N_y$ is odd and the other is even, then low crosstalk is
 impossible for the mode symmetry of the supermodes. Another condition for the low crosstalk
 intersection is that the two CCWs must be overlap at the center and perpendicular to
 each other.

 Although only a small number of $N_x$ and $N_y$ is discussed, one can straightly
 extend to the case of large values of $N$ ($N_x$ and $N_y$).
 However, there is a maximum value for $N$. Suppose the average linewidth of each mode
 is $\delta\omega^L$ and the
 bandwidth of the CCW is about $\Delta\omega$. The upper limit of $N$ is
 $N_{max}\sim\Delta\omega/\delta\omega$, and the maximum number of
 low crosstalk frequencies is
 $[N_{max}/2]$ ($[\cdot]$ means the integer part of a number).
 The bandwidth of the CCW can be tuned by changing the
 coupling of neighboring cavities.\cite{ccow:yariv,ccow:prl} And
 the linewidth of each mode can be
 tuned by the confinement of the cavities.\cite{supermode:dwq}
 For the  case of supermodes of CCW, the couple between cavities is much stronger
 than the case of continuous modes, therefore the bandwidth of $\Delta\omega$ is very large.
 For the structures
 demonstrated in this paper, the bandwidth (bandwidth to center frequency ratio $\delta\omega/\omega$)
 is about $10\%$, which is much larger than the bandwidth demonstrated
 before.\cite{cros2:SLan,cros1:StJohn,cros3:OL04,cros3:ptl05}

 In this paper, the reflection spectrum is not considered. Although the
 maximum normalized transmission power is only $85\%$ ($\omega_2$ in
 Fig. \ref{fig:Tb}), we belive that the reflection mainly originates from the
 coupling of the CCW and the incident source, but not from the existence of the
 intersection. The reflectance can be
 optimized using some tapered structure,\cite{taper:jlt,taper:ol} or using some novel optimization
 algorithm.\cite{cros3:ptl05,wannier}

 Finally, we want to point out that for the other eigenfrequencies
 (corresponding to the cases of $L=1,3,\cdots2m-1$ for the case of $N=2m$ or
 $L=1,3,\cdots2m+1$ for the case of $N=2m+1$),
 due to the symmetry of the mode
 profiles, the coupling of the two CCWs is strong. And this results in the supermodes split into
 more than one supermodes, and the crosstalk of them are all very high.
 In fact, the crosstalk are
 at the same level as the transmission power. Therefore, for these modes, the intersection
 structure performs just like a 1-to-3 power splitter, rather than a low crosstalk intersection.

 \section{Conclusions}
 In summary, we have investigated low crosstalk intersections of two
 CCWs that are composed by mono-mode cavities in a square lattice PC
 structures. The desired mode symmetry (orthogonal to the perpendicular CCW)
 is achieved by the combination of all the cavities but not a single cavity near
 the intersection. Our results show that for a cavity number of $N$,
 there are $[N/2]$ (the integer part of $N/2$) frequencies are low crosstalk.
 We analyzed the mode profiles using a coupled mode theory, and obtained the
 transmission and crosstalk spectra using FDTD method. We also
 obtained the electric field distributions at stable states of the
 low crosstalk modes, from which we analyzed the physical mechanism of low
 crosstalk.

 %%%%%%%%%%%%%%%%%%%%%%%%%%%%%%%%%% References and links %%%%%%%%%%%%%%%%%%%%%%%%%%%%%%%

 \end{document}